# Rescaling the dynamics of evaporating drops

C.Poulard[1], G.Guéna[1], A.M.Cazabat[1]*

A.Boudaoud[2], M.Ben Amar[2]

[1] Collège de France, 11 place Marcelin Berthelot, 75231 Paris Cedex 05
[2] Laboratoire de Physique Statistique de l'Ecole Normale Supérieure
24 rue Lhomond, 75231 Paris Cedex 05

- *Corresponding author

Abstract:

The dynamics of evaporation of wetting droplets has been investigated experimentally in an extended range of drop sizes, in order to provide trends relevant for a theoretical analysis. A model is proposed, which generalises Tanner's law, allowing us to smooth out the singularities both in dissipation and in evaporative flux at the moving contact line. A qualitative agreement is obtained, which represents a first step towards the solution of a very old, complex problem.

keywords
Moving contact line, evaporating drops.

**I. Introduction**

For practical reasons, the evaporation of drops, either free in aerosols [1-2] or deposited on fibres [3-5] or flat substrates [6-13] has been extensively studied in the past, and is still the subject of many investigations [14-19]. While the most recent studies deal with the structure of deposits from evaporating colloidal dispersions [20-22], the mere problem of the dynamics of evaporation of pure liquid drops on flat, smooth, horizontal substrates is not yet fully understood. The main difficulty is encountered in the complete wetting case, where the contact line moves freely on the substrate, either advancing or receding on it [23-25]. From the theoretical point of view, two singularities have to be coped with simultaneously, one associated with the well-known problem of a moving contact line, the second one with the specific behaviour of the evaporation flux at the edge of the drop. From the experimental point of view, the dynamics reveals to be rather sensitive to the surface energy, which is usually not the case in complete wetting. Actually, for non volatile wetting liquids, the dynamic properties depend only logarithmically on the spreading parameter [25].

The present paper brings a new piece towards a better understanding of the dynamics of evaporation of completely wetting drops of pure liquids on smooth substrates.

**Summary of our previous findings [17-19]**

In order to address the basic questions without too much interference of complex physico-chemical behaviour, we chose to study alkane drops evaporating on oxidised silicon wafers under normal atmosphere. When properly cleaned, the amorphous silica surface is hydrophilic, ensuring complete wetting, and the underlying silicon provides the high optical contrast required for interferometric measurements. A proper choice of alkanes, from hexane to nonane, allows us to vary the evaporation rate by a factor 35 while retaining the assumption of a diffusion-controlled, quasi stationary evaporation process, at least for the main part of the drop's life [21]. A consequence is that the laplacian $\Delta c$ of the concentration $c$ of the evaporating compound in the atmosphere is $\Delta c = 0$ [26].

This assumption is well obeyed in aerosols, and leads to simple laws for the radius (or the area) of the evaporating spheres: if $t_0$ is the time where the drop disappears, the radius scales with time $t$ as $\sqrt{t_0 - t}$ [27-28]. The square root dependence basically results from $\Delta c = 0$ and is therefore quite robust. For sessile wetting drops, the presence of a third phase and a contact line obviously makes the situation more complex. However, during the receding motion, the radius of the wetted spot is found experimentally to scale as $(t_0 - t)^y$, where y is close to 0.5, more precisely between 0.44 for nonane and 0.48 for hexane. The quality of the fit is excellent over the whole retraction.

Therefore, the dynamics of the radius is essentially defined by general trends, and the laws for sessile drops differ only by higher order terms from the "diffusive dynamics" of spherical ones.

In contrast, the contact angle is very sensitive to all the parameters involved: the volatility of the liquid, the surface energy of the substrate, and more insidiously the thermal conductivity of the phases and the velocity of the moving contact line. For non-volatile wetting liquids, the advancing dynamic contact angle depends on the contact line velocity and the receding one is zero, with a thin film being left on the substrate. In the present case, the receding contact angle does not vanish but typical values are small [17-19]. The experiments show that in the main part of the retraction, the contact angle decreases slowly, and scales acceptably as $(t_0 - t)^x$, where $x$ is positive and small. Combining the assumption of a diffusion-controlled, quasi stationary evaporation process, and the assumption that the drop is a spherical cap, one may predict that $x$ and $y$ are linked by $2y + x = 1$, which is actually well obeyed. Without surprise, $x$ depends on the difference between $y$ and the value 0.5 of the dynamics of free drops.

However, no prediction is available still for the precise value of $y$ (or $x$), and for the dependence of the contact angle with liquid volatility and substrate energy. Moreover, the behaviour of the contact angle, either at the beginning of the retraction, or at the very end of the drop life [19], differs significantly from the slow decrease mentioned above. Therefore, further experiments are required, in order to guide the theoretical analysis. From this point of view, sensitive parameters as contact angle and drop profiles are precious indicators, although difficult to work with.

**II. New experiments**

Two types of experiments have been performed:

In the first one, the experimental set-up is the same as in the previous papers [17-19]. The dynamics of wetting drops of volatile alkanes is followed under microscope, in normal atmosphere, with protections against air draft. The radius and contact angle are recorded versus elapsed time both during spreading (advancing motion) and retraction (receding motion). The data are conveniently plotted versus $t_0$-$t$, where $t_0$ is the time where the drop disappears. In contrast to the previous studies, the volume of the drop is varied significantly, and the crossover between spreading and retraction is investigated.

The study of the crossover has to be done carefully. The magnification needed for measuring the angle from equal-thickness interference fringes there prevents us to measure the radius simultaneously without manipulation of the objectives, which is out of question because it is enough to change the contact angle significantly. The radius is calculated from the visible part of the contact line, and the result is checked against an experimental curve (maximum radius versus initial volume of the drop), obtained independently and used as a "standard" (figure 1).

The aim of the second set-up is rather to spot behaviours which are not consistent with the various assumptions used in the analyses. A parallel monochromatic beam is sent normal to the drop. The silicon wafer is a plane mirror, and the reflected light is analysed (figure 2).
- part 1 of the beam does not meet the drop. We shall not discuss it further.
- part 2 is reflected by the drop surface.

- part 3 passes through the drop, is reflected on the silicon, then goes back.

Therefore, the drop acts as both a mirror (2) and a lens, which is crossed twice (3). The contact angle is small, a few degrees at the most. The two beams are received on a screen, directly or through a lens. It is possible to insert masks on the incident beam in order to illuminate a fraction of the drop, either at the centre, or in the vicinity of the edge. From the points of convergence of the beam (3) and from the size of the various spots on the screen, information on curvature of the drop, then the radius and contact angle is obtained.

If the drop is a spherical cap, the results are the same with or without masks, and the point of convergence of beam (3) on the axis is well defined (see figure 2).

If the drop is not a spherical cap, because of gravity, or because of any surface tension gradient, the results are different. If the drop is flatter, the lens with the mask on the centre has a shorter focal length, and conversely.

With the mask on the centre, the convergence of beam (3) is well defined and allows us to deduce $R/\theta$ in the vicinity of the contact line. In contrast, with a hole of radius $\rho$ = 2mm, the light is distributed on a small segment of the optical axis. The shortest length of convergence corresponds to the part of the drop located approximately at a distance $\rho$ from the centre and allows us to calculate $\rho/\theta_{(r=\rho)}$. The longest corresponds to the curvature at the centre, but the difference is less than 10% and will be ignored in the following.

The experiment provides a useful global picture of the behaviour of the evaporating drop, and information on the shape, which is quite important: as a matter of fact, the complete profile can be obtained from interference fringes only at the very end of the drop life.

**Summary of the results**

First, the radius $R$ of drops of nonane (figure 3a), octane (figure 3b), heptane (figure 3c) and hexane (3d) are plotted versus the time interval $t_0$-$t$ for different values of the initial drop volume, both in the advancing and receding motion. The range of volumes investigated is much larger than in the previous studies [17-19], therefore it becomes clear that the receding part of the various curves do not superimpose exactly, contrary to our previous conclusions. The locus of the extrema in the log-log plot is a straight line with slope ~ 0.64, the same for the four alkanes investigated.

In the second series of figures, the contact angles $\theta$ measured under microscope are plotted versus the time interval for two different volumes of heptane drops. As previously explained, measuring the contact angle of an evaporating drop is difficult. Moreover, the noise on the data is accentuated by the log-log representation (figure 4), because the angles are close to zero. The oscillations in the vicinity of the maximum extension of the drop are very reproducible and not due to any air draft, they are more visible on the linear plot (insert). As previously mentioned, there is a steep decrease of the contact angle during the spreading and the beginning of the retraction, then a range where the contact angle is almost constant, and where the relation $2y+x = 1$ is obeyed. Finally there is again a fast decrease at the end of the drop. The log-log curves for the two volumes are shifted in the main part, and tend to merge at the end.

The second set-up allows us to compare the results obtained with the different masks (figure 5). The straight lines on figures 5a and 5b have the slope $y$-$x$, which is expected for the ratio $\frac{R}{\theta}$ if the drop is a spherical cap.

The drop shape can be distorted by (i) gravity and (ii) surface tension gradients.

(i) As far as gravity is concerned, the maximum radius of the biggest drops is larger than the capillary length $a = \sqrt{\frac{\gamma}{\rho g}}$ ($a \approx 1.8$ mm), which means that these drops will be flatter than spherical. Let us give an estimate of the flattening of the drop. In the lubrication approximation, the static, three-dimensional drop profile is given by the equation:

$$-\left(\frac{d^2 h}{dr^2} + \frac{1}{r}\frac{dh}{dr}\right) = K_E - \frac{\rho g}{\gamma} h \qquad (1)$$

Here $\gamma$ is the surface tension, $\rho$ the density, $g$ the acceleration of gravity, $h$ the local thickness, and $K_E$ the curvature at the contact line, which is simply $\frac{2\theta}{R}$ for spherical caps. For flattened drops, the two radius of curvature are the same at the centre ($r = 0$), but no longer at the edge ($r = R$). The present experiment measures $\frac{R}{\theta}$ at the edge and $\left(-\frac{d^2 h}{dr^2}\right)^{-1}_{r=0} = \lim(r \to 0)\, r\left(\frac{dh}{dr}\right)^{-1}_{r \to 0}$ at the centre.

Solving equation (1) for a given contact angle allows us to calculate the various parameters as a function of the ratio $a/R$, and compare them with the ones of a spherical cap with same contact angle and radius. The value of the contact angle is chosen to be $\theta = 0.012$ rd, which is typical of octane. Then equation (1) is solved with $K_E = \frac{2\theta}{R}$.

On figure 6, the ratios of the curvatures, of the drop heights, and of the drop volumes have been plotted as a function of $a/R$ for $\theta = 0.012$ rd. The curvature is the most sensitive parameter, and the calculations are in good agreement with the measurements on figure 5. The fact that the lines with slope $y$-$x$ fit the data at the contact line even for flattened drops means that the volume of these drops is not yet significantly different from the ones of spherical caps with same $R$ and $\theta$, as can be seen also in the figure. It would no longer be the case for larger drops.

(ii) Surface tension gradients can be induced by evaporation. They are directed outwards if the edge of the drop is colder, which is the case during spreading [29-30], and then flatten the drop. They fade out when the drop has reached its maximum extension, and stays some time there at constant radius while the contact angle decreases with oscillations superimposed. No evidence of gradients during retraction can be inferred from the drop profile, because the mere gravity is enough to account for the observed shape.

The conclusion is that, for the alkanes considered, small drops, i.e., with radius of the order of the capillary length or less, are spherical caps, while larger ones are flatter. Including gravity allows us to account for the observations, without introducing any surface tension gradient (which does not mean that there is no gradient, only that they play as higher order terms). Moreover, the incidence of gravity on the dynamics seems to be weak, in the range of drop sizes investigated.

One must be aware that evaporation takes place during spreading as well, and that the volume of the drop when it starts to recede is much smaller than the initial volume. From the measured contact angle and radius, we can estimate the volume of the drop at the beginning of retraction, a value which is slightly overestimated if the drop is flatter than spherical, because we use the contact angle measured with the microscope. The volume of the drop at the maximum extension is approximately 35% of the initial volume for nonane, 30% for octane, 25% for heptane, 15% for hexane. Therefore we refrain to propose models for the curves on figure 1.

**III. Theoretical analysis: rescaling**

The starting point for the evolution of the drop is the local conservation equation:

$$\frac{\partial h}{\partial t} + \nabla(hU) = -J(r) \qquad (2)$$

Here, $h$ is the local thickness, $r$ the distance to drop axis, $\eta$ the viscosity of the alkane, $U$ the velocity averaged over the thickness.

$$U(h,t) = \frac{h^2}{3\eta}\nabla(\gamma \Delta h - \rho g h + \Pi(h)) + \frac{h}{2\eta}\nabla\gamma \qquad (3)$$

$J(r)$ is the evaporation rate per unit area (of the surface of the substrate), and $\Pi(h)$ the disjoining pressure. Ignoring surface tension gradients, which is supported by the previous discussion, one gets:

$$U(h,t) = \frac{h^2\gamma}{3\eta}\nabla(\Delta h - \frac{h}{a^2} + \frac{\Pi(h)}{\gamma}) \qquad (4)$$

where $a = \sqrt{\frac{\gamma}{\rho g}}$ is the capillary length.

With the assumption of a diffusion-controlled, quasi stationary evaporation process, and considering that the contact angles are very small, the evaporation rate can be written as [14, 20-21]:

$$J(r) = \frac{j_0}{R\sqrt{1-\left(\frac{r}{R}\right)^2}} \qquad (5)$$

where $R$ is again the radius of the drop. This quantity diverges at the edge of the drop and has to be regularised there.

Equation (2) is conveniently rescaled using characteristic lengths and times. A characteristic length should be some radius, which from the experimental study can be chosen as the maximum radius of the drop $R_0$. A characteristic thickness in the lubrication approximation is $R_0\theta_0$, where $\theta_0$ is logically the angle corresponding to $R_0$ taken at the beginning of the retraction. The characteristic time is then $\dfrac{R_0^2 \theta_0}{j_0}$.

Let us now proceed with the equation. Rescaling introduces several dimensionless quantities:

$$C = \frac{3\eta j_0}{\gamma R_0 \theta_0^4} \quad \text{is a capillary number}$$

The influence of the gravity is contained in the Bond number

$$\alpha = \frac{R_0^2}{a^2}$$

The disjoining pressure will play in the regularisation at the edge of the drop. Let us assume pure van der Waals interaction, and let H be the absolute value of the Hamaker constant. The disjoining pressure is positive in complete wetting and can be written as:

$$\Pi(h) = \frac{H}{6\pi h^3}$$

The second dimensionless quantity is a "van der Waals" number

$$A = \frac{H}{6\pi \gamma R_0^2 \theta_0^4}$$

The equation becomes, keeping the same symbols for the dimensionless variables:

$$\frac{\partial h}{\partial t} + \frac{1}{C}\nabla \cdot \left( h^3 \nabla \left( \Delta h - \alpha h + \frac{A}{h^3} \right) \right) = -\frac{1}{R\sqrt{1-\left(\dfrac{r}{R}\right)^2}} \qquad (6)$$

$A$ is very small, but $C$ and $\alpha$ are of order 1. For example $C = 0.16$ and $\alpha = 4$ for a 3µl drop of heptane, where $R_0$ = 3.6 mm. The same orders of magnitude are obtained with the other alkanes.

**Rescaling: check against experiments**

The relevance of the rescaling can be checked easily on the experimental log-log plots, because it corresponds to a mere translation such that the curves coincide at the maximum extension.

For a given alkane, the curves for the radius merge very well if they are translated parallel to the line with slope 0.64. A true rescaling has been done for heptane, where precise measurements of the angle are available, and works very well, see figure 7.

For the angle, the rescaling works also, except at the very end of the drop, where the decrease of the angle is faster, see figure 8.

Note that $R_0$ and $\theta_0$ are not independent quantities. From direct measurements in heptane, the trend is that $\theta_0 \propto R_0^Z$, with z ≈ - 0.4. The shifts needed to superimpose the curves for the radius in the log-log plot gives the best estimate. Here the vertical shift is $Log R_0$, the horizontal one $Log \theta_0 R_0^2$, the slope is - 0.63, which gives z = - 0.45 (see figure 3)

In conclusion, the rescaling works well, except at the very end of the drop. It is clear from figure 4 that in that range, the curves for the contact angle have a tendency to merge before scaling but not after, which deserves further discussion.

One obvious explanation for the faster decrease of the angle at the end of the drop is that the velocity becomes large there, which in a receding motion reduces the angle. This is true, but this effect is already taken into account in the hydrodynamic term and therefore properly rescaled.

On the other hand, one must be aware that the rescaling in time depends directly on the structure of the evaporation term, i.e.,

$$J(r) \propto \frac{j_0}{R} \qquad (7)$$

which means that the rate of change of the volume $V$ of a drop with small contact angle is proportional to the radius $R$

$$\frac{dV}{dt} \propto R \qquad (8)$$

If the drop is a spherical cap, and if power laws for radius and angle versus $(t_0 - t)$ are still observed, the corresponding exponents y and x have to be linked by $2y+x = 1$.

Experiments show that the drops keep a perfect spherical cap shape till the very end [19]. In that range, the fit with a power law is excellent for the radius with no appreciable change in y (y ~ 0.5). The uncertainty is larger for the angle (figure 4), however the data are acceptably fitted by a power law where x ~ 0.4-0.5, in sharp disagreement with the relation $2y+x = 1$. This suggests that equation (7) is probably not valid at the end of the drop life.

The change in evaporation rate may have various origins. As previously noted, the assumption of a diffusion-controlled, quasi stationary evaporation process is probably no longer acceptable at the end of the drop, because the interface velocity is large, and convection should take place [18]. Another possibility might be that the disjoining pressure plays over a significant fraction of the drop, but this is discarded by experiment, because several interference fringes are still observed in that range. The disjoining pressure is completely negligible above the first fringe. Moreover, previous calculations showed that the influence of van der Waals interaction on evaporation rate is significant only for film thickness of the order of nanometer [18].

Whatever the cause of the change in the evaporation rate, the rescaling has to be done accordingly. This could possibly explain why the contact angle seems to depend only weakly on the drop volume at the very end of the drop's life.

We shall restrict the discussion to the range where the rescaling works well.

**IV. Theoretical analysis: regularisation**

The rescaled equation must now be regularised, in order to obtain the dynamics of the moving contact line. Let x be the (rescaled) distance to the edge. In the frame of the contact line, which moves with velocity $\frac{dR}{dt}$, and for x << 1, the equation can be written as:

$$\frac{dR}{dt}\frac{\partial h}{\partial x} + \frac{1}{C}\frac{\partial}{\partial x}\cdot\left(h^3 \frac{\partial}{\partial x}\left(\frac{\partial^2}{\partial x^2}h + \frac{A}{h^3}\right)\right) = -\frac{1}{\sqrt{2Rx}} \qquad (9)$$

A way to remove the divergences is first to calculate the profile of the drop edge as a balance between van der Waals interaction and curvature. The procedure was proposed by de Gennes and Hervet [31] for the edge of the precursor film in complete wetting. The profile at the edge is given by:

$$\frac{\partial^2}{\partial x^2}h + \frac{A}{h^3} = 0 \qquad \text{which leads to} \qquad h = A^{1/4}\sqrt{2x}$$

The crossover towards the main drop with contact angle $\theta$ takes place over a distance $\ell \approx \frac{A^{1/2}}{\theta^2}$.

The angle at the very edge is now 90°, which also removes the divergence in the evaporation flux. As a matter of fact, the behaviour of the flux $Q$ per unit area (of the drop surface) close to a wedge of contact angle $\varphi$ is [20-21, 32]:

$$Q \propto (R-r)^{-\lambda}$$

with

$$\lambda = \frac{\pi - 2\varphi}{2\pi - 2\varphi}$$

$\lambda$ is 0.5 for small angles, but tends to zero if $\varphi \to \frac{\pi}{2}$, which means that the flux is finite. Therefore, a maximum value of the flux $Q$ will be obtained, which might be estimated as $\frac{1}{\sqrt{R\ell}} = \frac{\theta}{2^{1/2} A^{1/4} R^{1/2}}$.

At a distance $\ell$ from the edge, the slopes are small so that $J = Q$ which gives a maximum value for the right hand side of equation (9).

One may argue that, in the reality, the wetting drop recedes over a flat film, which is thin enough not to evaporate. If d is the corresponding thickness, the crossover between main drop and film scales as $\dfrac{d}{\theta^2}$ [25]. This will lead to similar results, because the thickness d of the precursor is again controlled by the disjoining pressure. Another regularisation is to make the evaporation flux saturate at a distance $\ell = \dfrac{D}{v_{th}}$ from the periphery of the drop ($D$ is the diffusivity of the vapour in air and $v_{th}$ is a typical thermal velocity) as deduced from a generalised boundary condition at the liquid/gas interface derived in [33]. This leads again to similar results.

Evaluating the terms of equation (9) at a distance $\ell$ from the edge leads to

$$\frac{dR}{dt} - \frac{1}{C}\theta^3 = -\frac{1}{2^{1/2} A^{1/4} \sqrt{R}} \qquad (10)$$

up to numerical factors of order one. This equation is the generalisation of Tanner's law [34] accounting for evaporation.

In the range where the drops are spherical caps, volume conservation reads:

$$3R\theta \frac{dR}{dt} + R^2 \frac{d\theta}{dt} = -8 \qquad (11)$$

the system can be solved numerically.

The derivation of the mobility law for the contact lines is valid as long as $\ell < R$, a condition which is broken at the end of the drop life.

**Check against experiments**

The relevance of the regularisation procedure can be indirectly checked on equation (10). At the maximum extension, the rescaled $R$ and $\theta$ are equal to 1, and the velocity $\dfrac{dR}{dt} = 0$. Therefore,

$$C = 2^{1/2} A^{1/4} \qquad (12)$$

Coming back to the usual variables, one finds a relation between $\theta_0$ and $R_0$

$$\theta_0^3 = \frac{3\eta j_0}{\gamma}\left(\frac{4H}{6\pi\gamma}\right)^{-1/4}\frac{1}{\sqrt{R_0}} \qquad (13)$$

The orders of magnitude are correct, but the relation between angle and radius at the maximum differs significantly from the experimental power law $\theta_0 \propto R_0^{-0.45}$. That could be expected, because the regularisation has been done using an oversimplified procedure. However, the mere fact to be able to find a relation between maximum radius and angle is already a significant step towards the complete description of the evaporating wedge.

In the same way, the coupled equations (10) and (11) can be solved numerically, taking the maximum extension as the initial condition, with $R$ = 1, $\theta$ = 1, and $\frac{dR}{dt} = 0$, which implies that (12) is valid. Therefore the only free parameter is C. Taking the experimental value C = 0.16 one obtains the curves plotted on figure 9. The agreement is qualitatively excellent. Quantitatively, the decrease in angle is overestimated with respect to the experiment, and the results are very sensitive to the value of C, which is not fully plausible. More seriously, the angle vanishes before the radius and diverges towards $-\infty$. This behaviour is reminiscent of the inadequacy of the wedge model in partial wetting at large receding velocities [35] . How to define $t_0$ precisely is therefore not obvious. In fact, a singularity shows up in the calculation at a time, which seems to be a good candidate. The log-log plots on figure 10 have been obtained in this way.

**V. Conclusion**

It is clear that the present theory picks up a large part of the physics of the problem, and represents a significant step towards its complete understanding. It accounts well for the experimental power laws obtained for the radius of receding, evaporating drops. It also provides the first attempt to predict the value of the contact angle in a dynamic situation, and a generalisation of Tanner's law in the case of evaporating liquids. It has to be improved at some places. Noticeably, the changes in angle and the sensitivity of the solution to the value of C, i.e., to the hydrodynamic term, have to be smoothed out, which means that a real profile has to be introduced in the equations, and not only the slope at the edge. However, it provides the first complete and plausible description of the very old problem of evaporation of drops on solid substrates.

**Acknowledgements**

We gratefully acknowledge M.Brenner, H.Stone and M.Betterton, who gave access to unpublished work, and Elie Raphaël for enlightening discussions. Sergueï Mechkkov and Baptiste Mangeney took part in some of the experiments.

**References**

[1] N. Roth A. Frohn. *Dynamics of droplets.* Springer edition, 2000.
[2] N.Roth, K. Anders, and A. Frohn. Size and evaporation rate measurements of optically levitated droplets. In *Proc. 3rd Int Congr. on Optical Particle Sizing*, pp 371–377, 1993.
[3] B. Topley and R. Whytlaw-Gray. *Phil. Mag.*, **4** :873, 1927.
[4] H.G. Houghton. *Physics*, **4** :419, 1933.
[5] W.E. Ranz and W.R. Marshall. *Chem. Eng. Progress*, **48**(141) :173, 1952.
[6] H.W. Morse. *Proc. Amer. Acad. Sciences*, **45** :363, 1910.
[7] P.Ehrhard, S.H.Davis, J. Fluid Mech, **229**, 365 (1991)

[8] D.M.Anderson, S.H.Davis, Phys.Fluids **7**, 248 (1995)

[9] L.M.Hocking, Phys.Fluids **7**, 2950 (1995)

[10] C.Bourguès-Monnier, M.E.R Shanahan, Langmuir **11**, 2820 (1995)

[11] M.E.R Shanahan, Langmuir **11**, 1041 (1995)

[12] K.S.Birdi, D.T.Vu, A.Winter, J.Phys.Chem. **93**, 3702 (1989)

[13] M.Betterton, M.Brenner, H.Stone, private communication

[14] R.D.Deegan, Phys.Rev.E **61**, 475 (2000)

[15] S.J.S.Morris, J. Fluid Mech, **432** (2001)

[16] H.Hu, R.G.Larson, J.Phys.Chem.B, **106**, 1334 (2002)

[17] M.Cachile, O.Bénichou, A.M.Cazabat, Langmuir **18**, 7985 (2002)

[18] M.Cachile, O.Bénichou, C.Poulard, A.M.Cazabat, Langmuir **18**, 8070 (2002)

[19] C.Poulard, O.Bénichou, A.M.Cazabat, Langmuir **19**, 8828 (2003)

[20] R.D.Deegan, O.Bakajin, T.F.Dupont, G.Huber, S.R.Nagel, T.A.Witten, Nature (London) **389**, 827 (1997)

[21] R.D.Deegan, O.Bakajin, T.F.Dupont, G.Huber, S.R.Nagel, T.A.Witten, Phys.Rev.E **62**, 757 (2000)

[22] F.Parisse, C.Allain, Langmuir **13**, 3598 (1996)

[23] T.D.Blake, AIChE Spring meeting (New Orleans, LA, 1988) Paper I.a

[24] Y.Pomeau, C.R.Acad.Sci. Paris, **328**, série IIb, p.411 (2000)

[25] P.G. de Gennes, Wetting : statics and dynamics, *Rev. Mod. Phys*. **57**, V3, 827 (1985)

[26] J.C. Maxwell. Diffusion, collected scientific papers. *Encyclopedia Britannica, Cambridge*, 1877.

[27] I. Langmuir. Evaporation of small spheres. Physical review, **12** :368, 1918.

[28] N.A. Fuchs. *Evaporation and growth in gaseous media*. R.S. Bradley, Pergamon press edition, 1959.

[29] C.Redon, F.Brochard-Wyart, F.Rondelez, J.Phys.II, **2**, 580 (1992)

[30] O.Bénichou, M.Cachile, A.M.Cazabat, C.Poulard, M.P.Valignat, F.Vandenbrouck, D.Van Effenterre, Adv.Colloid Int.Sci. **100-102**, 381 (2003)

[31] P.G. de Gennes, H.Hervet, *"Dynamique du mouillage : films précurseurs sur solide sec"* C. R. Acad .Sci. Paris**, 299**, 499 (1984)

[32] J.D.Jackson, *"Classical Electrodynamics"* 2[nd] Ed. Wiley. New-York (1975)

[33] E.Sultan, A.Boudaoud, M.Ben Amar *"Evaporation of a thin film: diffusion of the vapour and Marangoni instabilities"* submitted J.Fluid Mech. (2005)

[34] L.H.Tanner, J.Phys.D **12**, 1478 (1979)

[35] R.G.Cox, J.Fluid Mech. **168**, p.169 (1986)

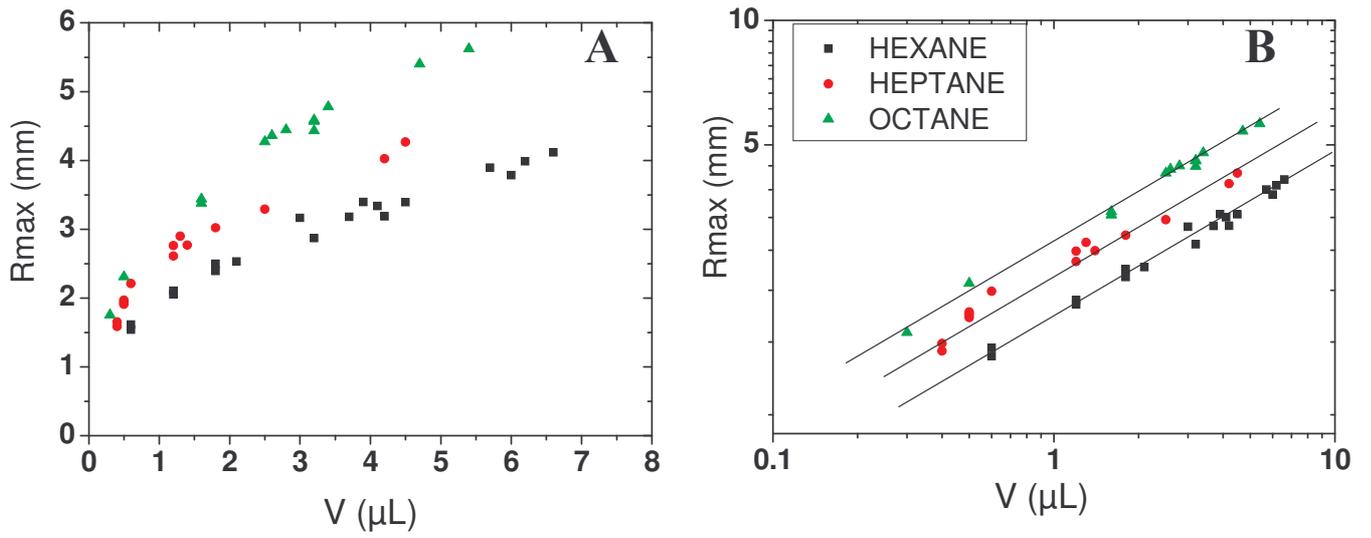

Figure 1 : (A) : Linear plot of maximum radii vs. initial volume for different alkanes. (B) : log-log representation of (A), the lines with slope (~0.4) are guides for the eyes.

**(A)**

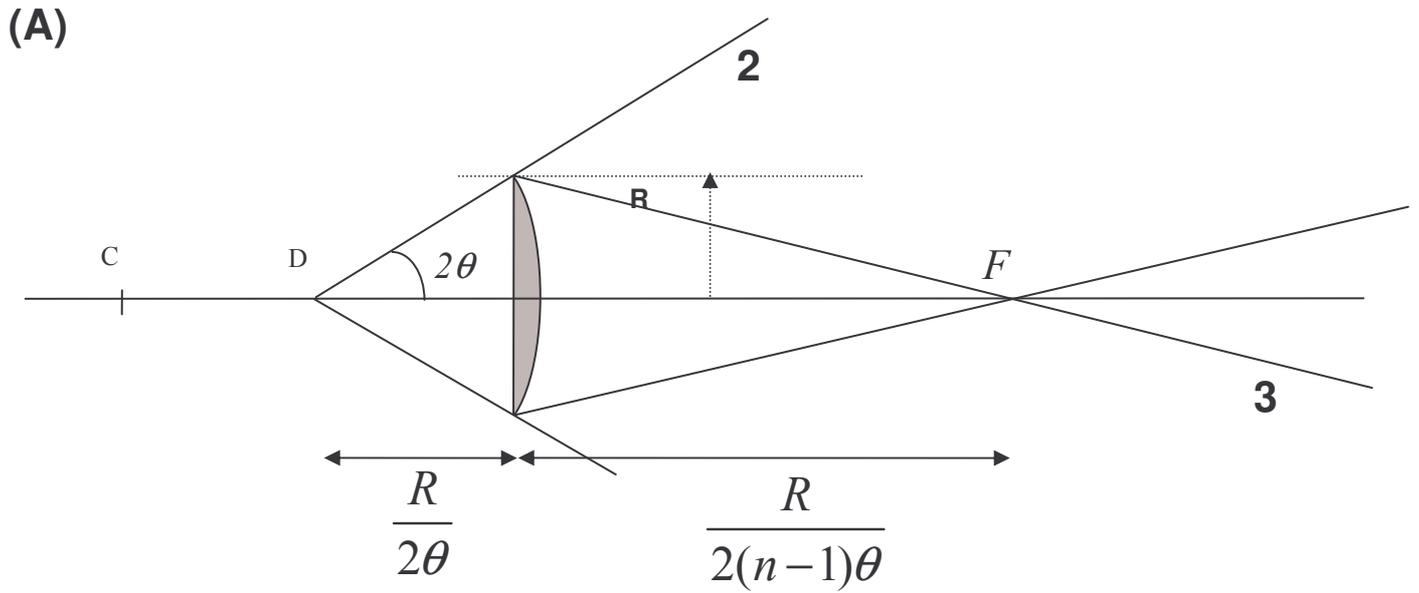

**(B)**

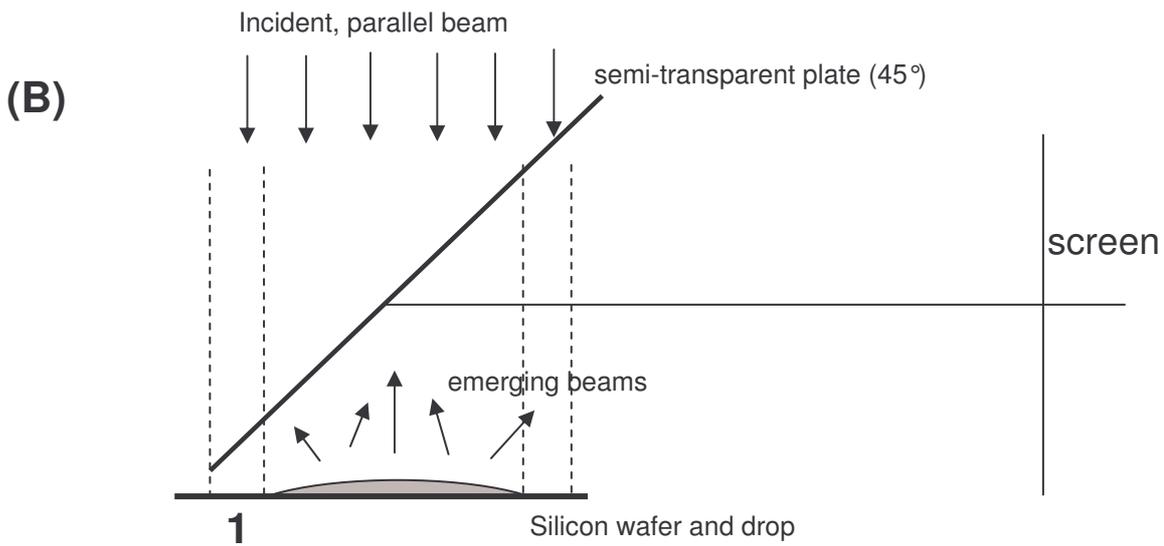

Figure 2: Schematic view of the optical setup (not to scale, the angles are in fact very small)
(A) definition of the emerging beams: C is the geometrical centre of the drop when it is a spherical cap, it is located at $R/\theta$ from the substrate. The emerging beam (2) is reflected on the drop surface and seems to come from point D, located at $R/2\theta$. The beam (3) has crossed the drop twice and converges at point F on the axis.
(B) Geometry of the setup.

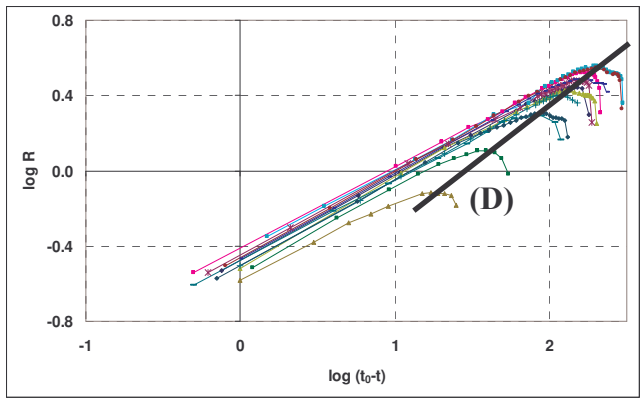

(a)

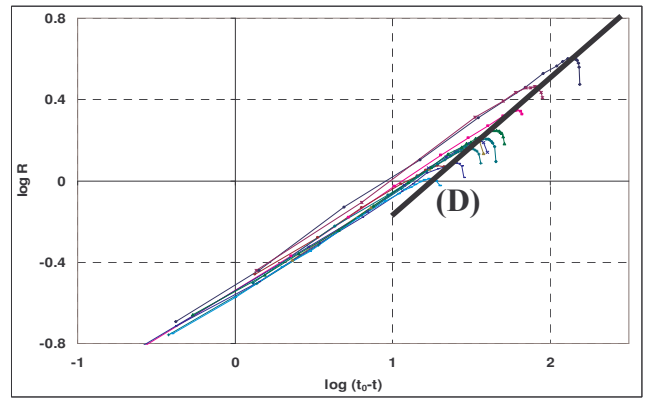

(b)

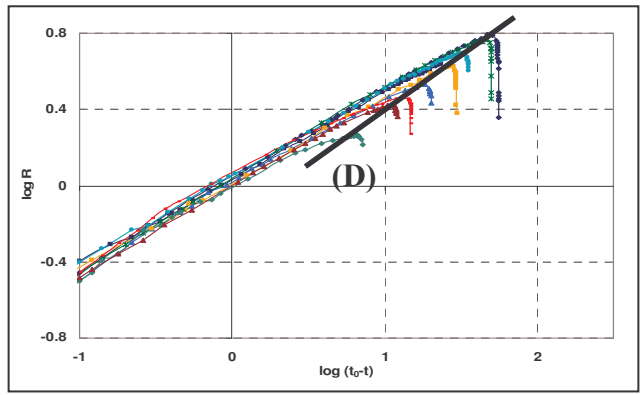

(c)

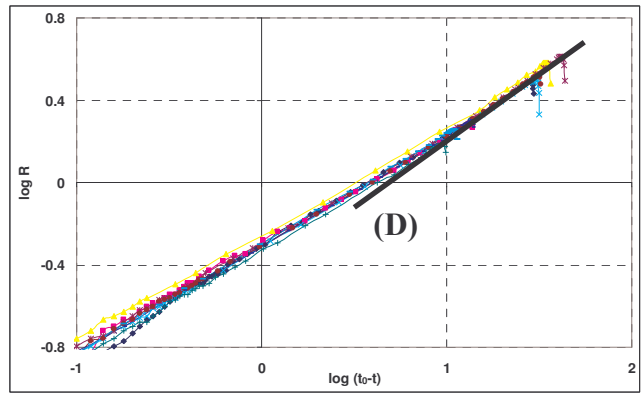

(d)

Figure 3 : log-log plot of the radius $R$ vs. time $(t_0-t)$ for different initial volume of alkanes droplets. The line (D) passes by maximum radii for each volume
(a) Nonane : The measured slope of (D) is 0.64.   (b) Octane : The measured slope of (D) is 0.64
(c) Heptane : The measured slope of (D) is 0.63.   (d) Hexane : The measured slope of (D) is 0.63.

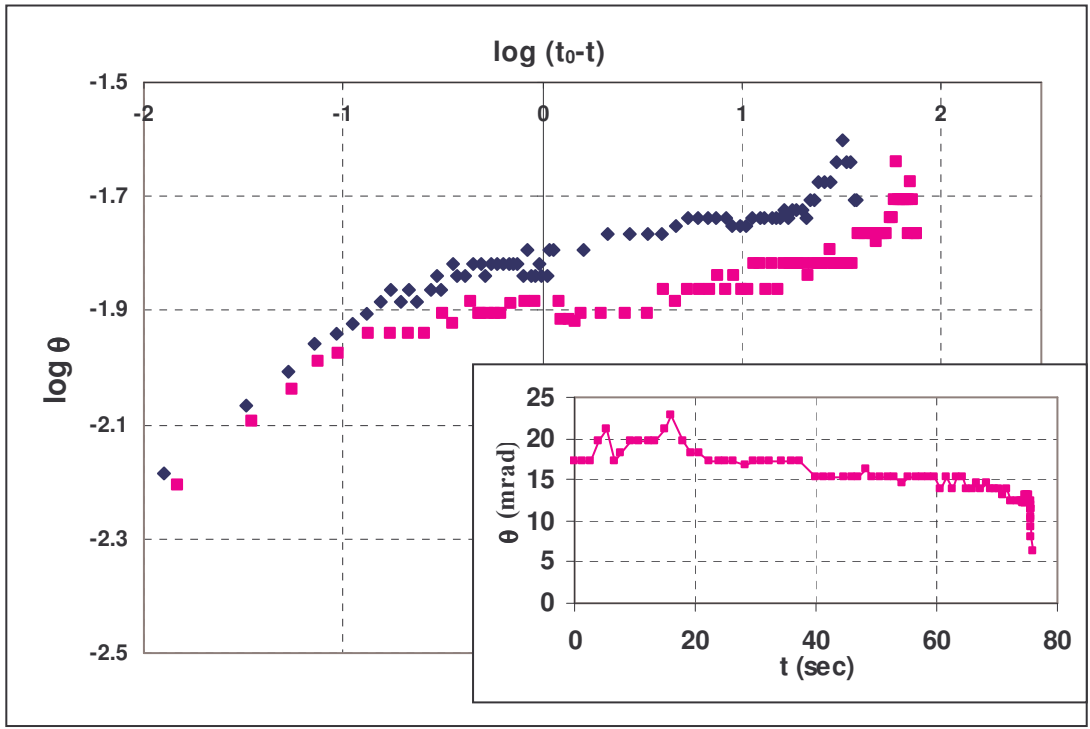

Figure 4 : log-log plot of the contact angle $\theta$ vs. time $(t_0-t)$ for different initial volumes of heptane droplets : (diamonds): $3\mu l$; (squares):$10\mu l$. Insert : linear representation of the contact angle vs. time ($3\mu l$).

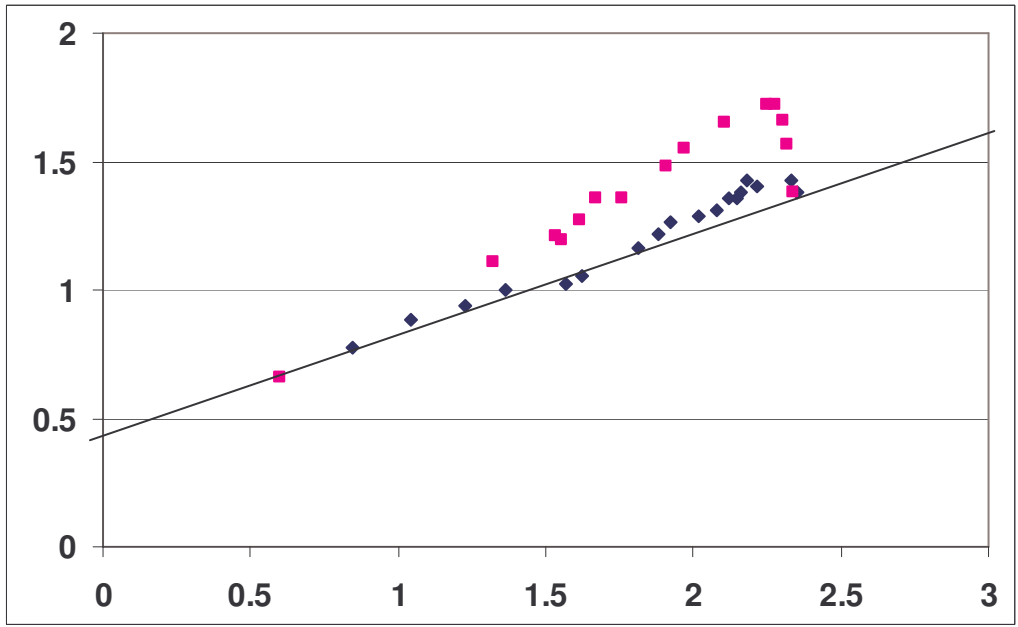

Figure 5a : log-log plot of radius of curvature *(cm)* versus $t_0$-$t$ *(s)* for a drop of octane of initial volume *4µl*, and maximum radius *5 mm*.

The diamonds give $\dfrac{R}{\theta}$ , which is one the radius of curvature at the contact line, the squares give $\left(-\dfrac{d^2h}{dr^2}\right)^{-1}_{r=0}$ which is the radius of curvature at the centre.

The slope of the straight line is $y - x \approx 0.4$

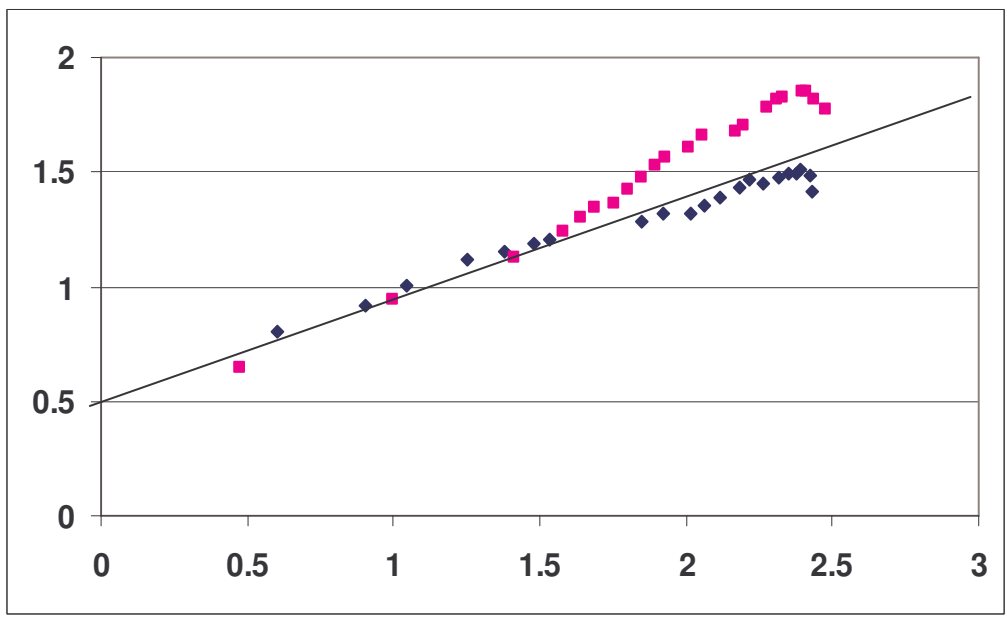

Figure 5b : log-log plot of radius of curvature *(cm)* versus $t_0$-$t$ *(s)* for a drop of octane of initial volume *7µl*, and maximum radius *6.5 mm*.

The diamonds give $\dfrac{R}{\theta}$ at the contact line, the squares give $\left(-\dfrac{d^2h}{dr^2}\right)^{-1}_{r=0}$ at the centre.

The slope of the straight line is $y - x \approx 0.4$

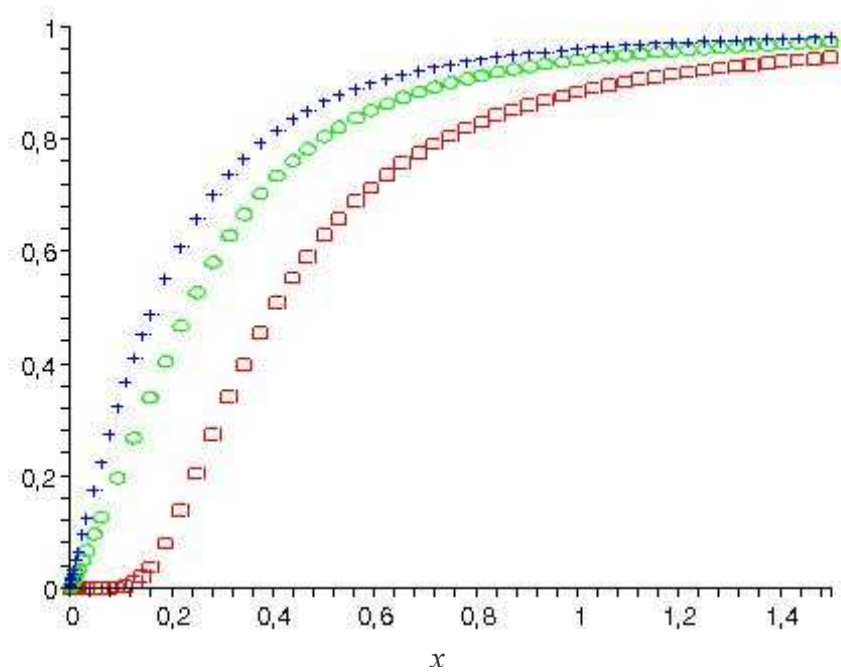

Figure 6 : Comparison between the parameters of the flattened drop and the ones of a spherical cap with same radius and contact angle $\theta$ = 0.012 rd, as a function of $x = a/R$.
Crosses: ratio of the volumes
Circles: ratio of the heights

Squares: ratio of the curvatures in a plane of symmetry of the drop: $\left(-\dfrac{d^2 h}{dr^2}\right)_{r=0}$ (curvature at the center) and $\dfrac{\theta}{R}$ (curvature at the edge).

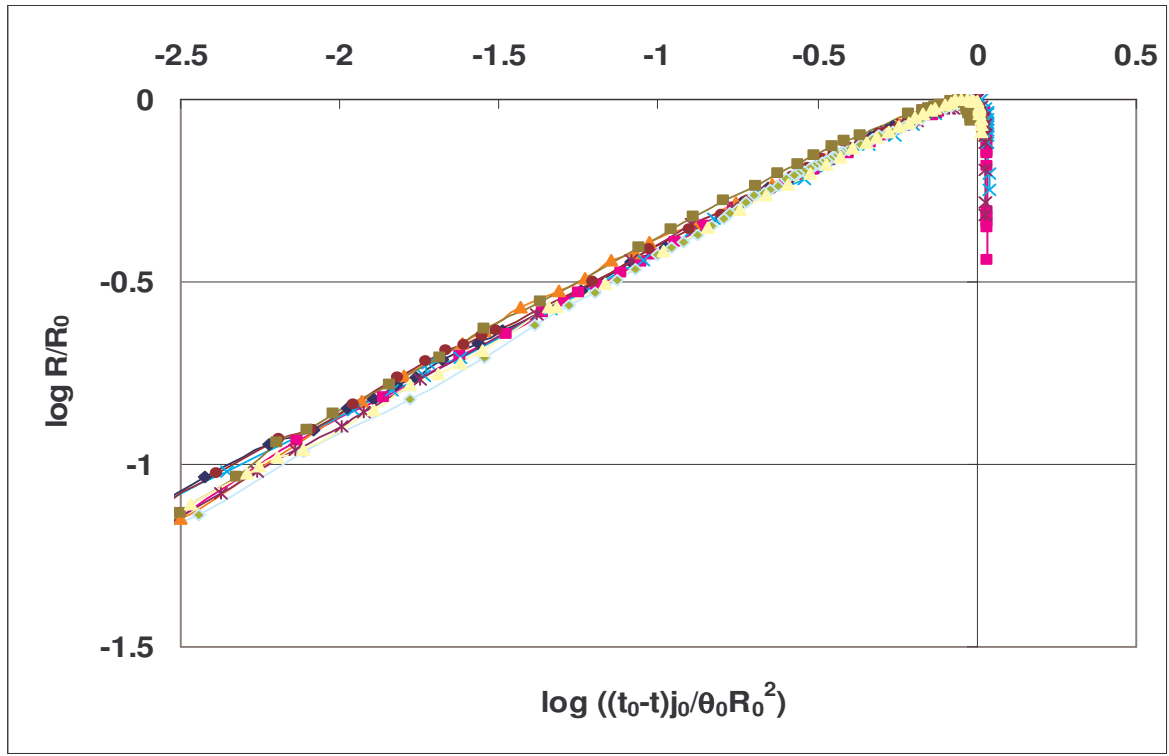

Figure 7 : log-log plot of the rescaled radius vs rescaled time for different initial volume of heptane droplets.

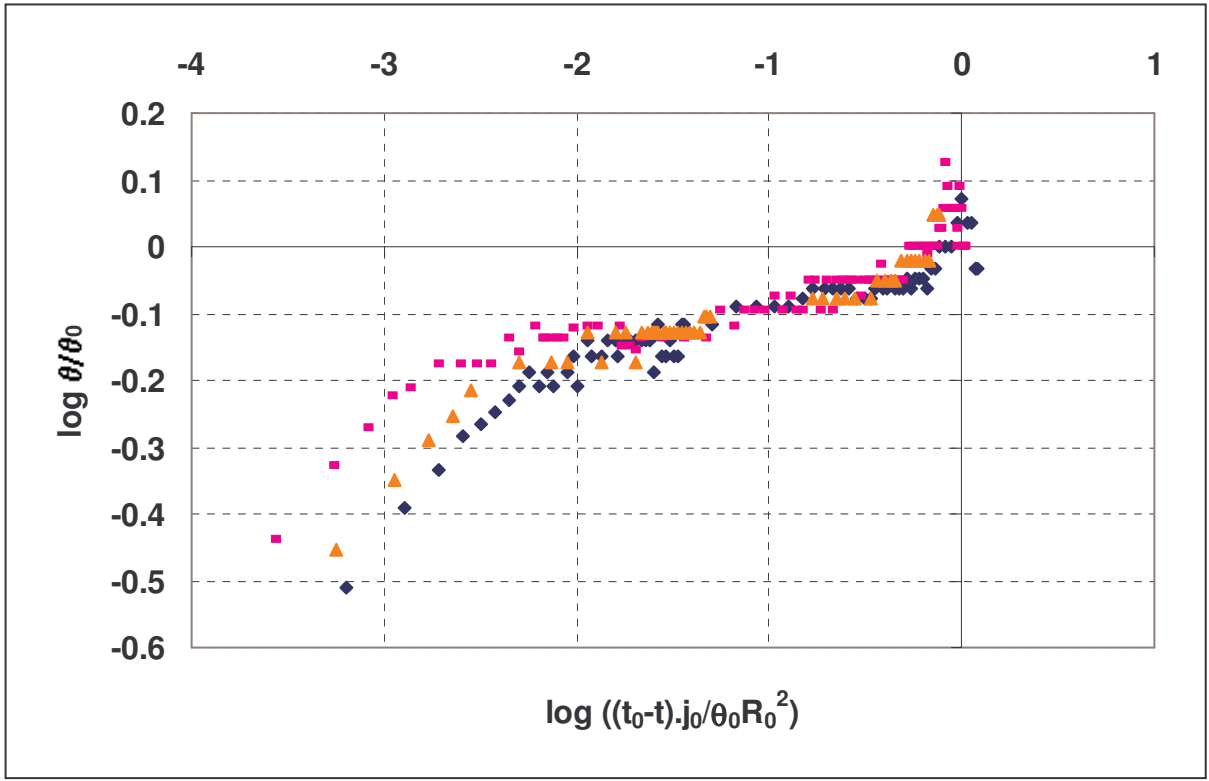

Figure 8 : log-log plot of the rescaled contact angle vs. rescaled time for different initial volumes of heptane droplets : (diamond) : 3µl; (triangle) :5µl; (square):10µl.

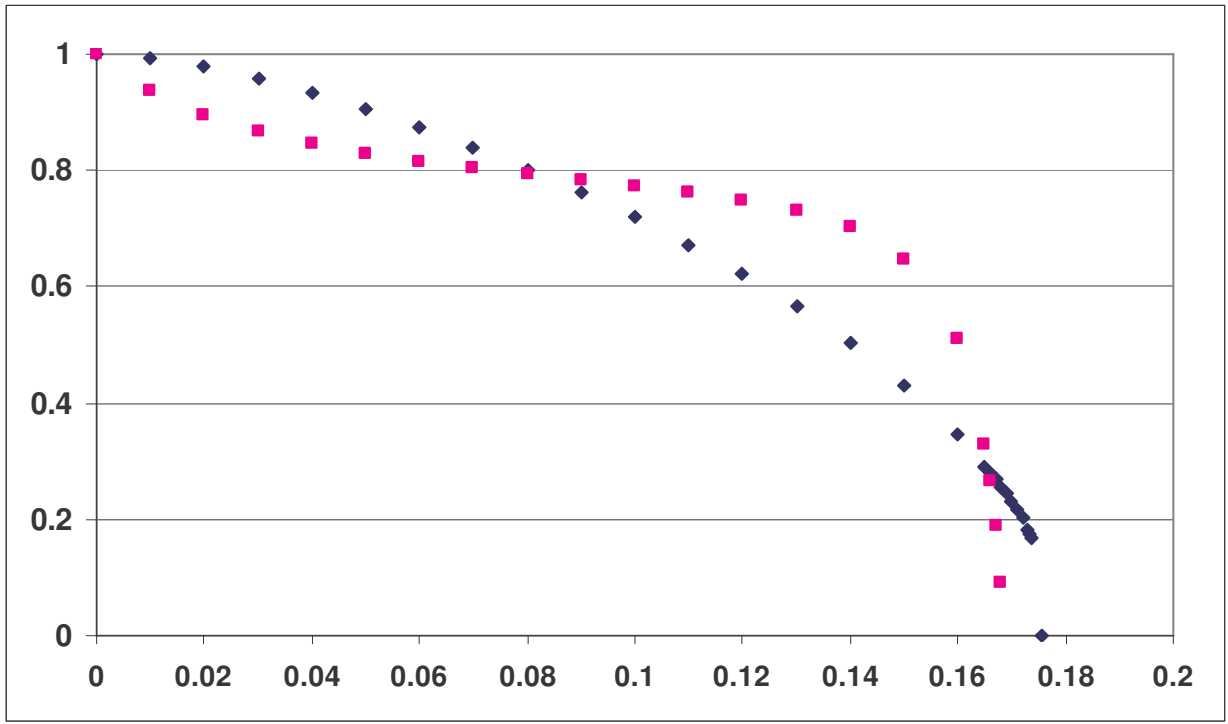

Figure 9 : Calculated rescaled contact angle $\theta$ (squares) and radius $R$ (diamonds) versus rescaled time $t$.

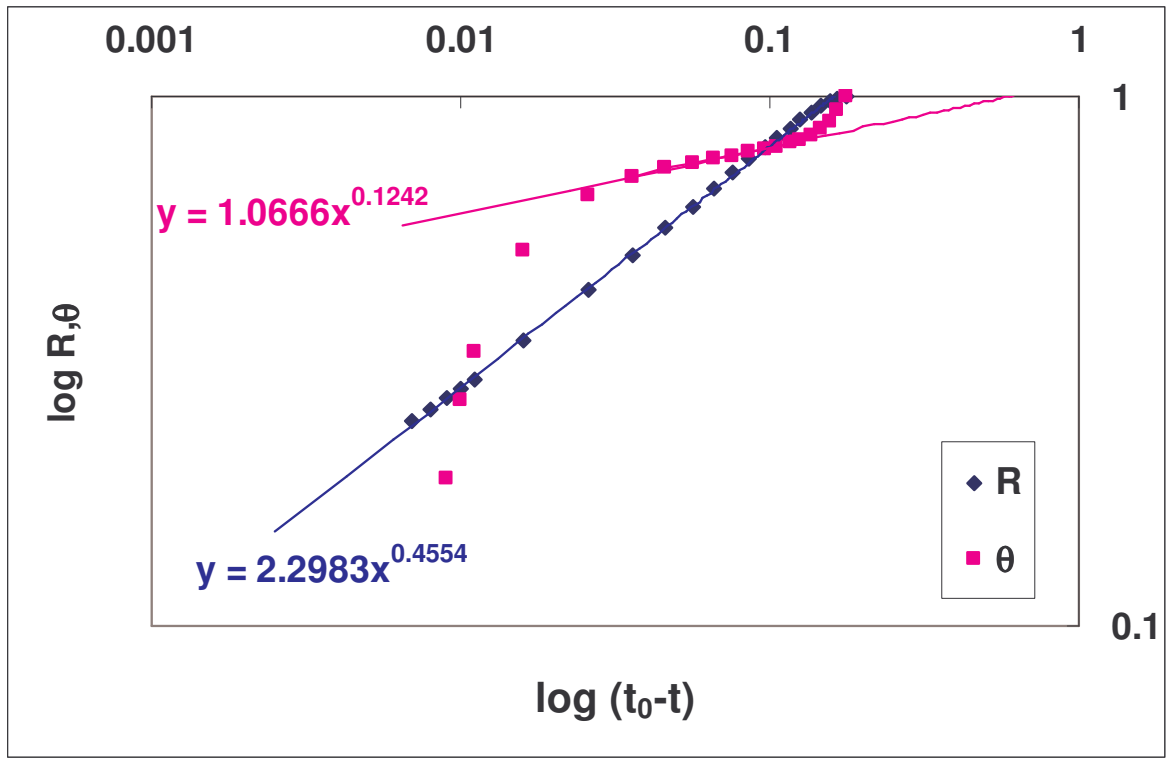

Figure 10 : log-log plot of the rescaled contact angle $\theta$ and radius $R$ vs. rescaled time $(t_0-t)$